# Unusual dynamics of Fe atoms in chromium matrix

S. M. Dubiel[*], J. Żukrowski, and J. Cieślak

Faculty of Physics and Computer Science, AGH University of Science and Technology, al. A. Mickiewicza 30, 30-059 Kraków, Poland

$^{57}$Fe site Mössbauer Spectroscopy (MS) was used to investigate a dynamics of $^{57}$Fe atoms embedded into chromium lattice as impurities. From the Mössbauer spectra recorded in the temperature range of 80 to 350 K, a temperature dependence of the Lamb-Mössbauer factor, $f$, was determined. The latter revealed an unusual dynamics of $^{57}$Fe atoms viz. harmonic mode below $T \approx 145$ K with a characteristic effective Debye temperature, $\Theta_{eff} = 185$ K and anharmonic one above $T \approx 145$ K. The latter mode exists in two clearly separated temperature intervals with slightly different $\Theta_{eff}$ – values viz. (i) ~156 K for ~145 K $\leq T \leq$ ~240 K and the record-high anharmonic coefficient $\varepsilon = 18 \cdot 10^{-4}$ K$^{-1}$ and (ii) ~152 K for $T \geq$ ~240 K and $\varepsilon = 13.6 \cdot 10^{-4}$ K$^{-1}$. Based on the Visscher's theory, the record-low values of relative binding force constants for Fe atoms were determined as 0.0945, 0.0673 and 0.0634, respectively. It is suggested that the unusual dynamics observed in this study might be related to the underlying spin- and charge- and strain-density waves of chromium.

Mössbauer Spectroscopy (MS) can be readily used to study a lattice dynamics through two different quantities: (i) the second order Doppler-shift, *SOD*, which is measurable via the center shift, *CS*, and (ii) the Lamb-Mössbauer factor or recoil-free fraction, *f*. It must be, however, remembered that the two quantities give information on the dynamics – the former via the mean-squared velocity and the latter via the mean-squared displacement, $<x^2>$, of the emitting atoms. Consequently, values of the Debye temperature, $\Theta_D$ as determined from *CS* and *f*, are different. For example, in metallic iron $\Theta_D = 421 \pm 30$ K was found from *SOD* against $\Theta_D = 358 \pm 18$ K from *f* [1]. MS is especially useful in investigations of the dynamics of impurities embedded in metallic matrices. The Debye temperature, which can be determined in such experiment is known as the effective Debye temperature, $\Theta_{eff}$, and it can give information about the binding force between the impurity and the host atom [2-5]. This kind of experiment can also give useful information on anharmonic effects that may strongly depend on the host matrix [2,5].

In this study we will use the temperature dependence of $f = exp(-\kappa^2 <x^2>)$, to study the dynamics of $^{57}$Fe atoms embedded as impurities (concentration < 0.1 at%) into Cr matrix, $\kappa$ being the wave number for the $\gamma$- rays. Using the Debye model for the lattice dynamics, *f* can be expressed as $f = exp(-2W)$, where for $T > \Theta_D/2$,

$$2W = \frac{6E_R T}{k\Theta_D^2} \quad (1)$$

$E_R$ being the recoil energy of an unbound emitting nucleus (equal to 22.6 K for $^{57}$Fe) and *k* is the Boltzmann constant. Equation (1) means that for high enough temperatures ln*f* is a linear function of *T*. Maradudin and Flinn have calculated the effect of anharmonic binding on *f* in the case of a pure crystal [6]. According to their calculations,

$$\ln f = -\frac{6E_R T}{k\Theta_D^2}\left(1 + \varepsilon T + \delta T^2 + ...\right) \qquad (2)$$

Here $\delta$ should be much smaller than $\varepsilon$ which can be then termed as the anharmonic coefficient.

The sample for the present investigation was prepared by melting elemental chromium (3N-purity) and iron enriched to ~95% in the $^{57}$Fe isotope mixed in a right proportion to give ≤ 0.1 at% Fe. The melting process, that was carried out in an arc furnace under a protective atmosphere of a pure argon, was repeated 3 times to ensure a better homogeneity. For Mössbauer-effect measurements the ingot was filed to powder with a diamond file. $^{57}$Fe-site spectra, whose typical examples recorded at different temperatures in the range between 80 and 350 K, are shown in Fig. 1.

To obtain the *f*-values and to take into account the absorber effective thickness the spectra were fitted with a single component with the main goal to best reproduce their shape. For this purpose a procedure based on the transmission integral method was used. It was assumed that *f*, hence the spectral area, is proportional to the effective thickness, *t*, which was a fit parameter. Hence $t/t_o = f/f_o$, where $t_o$ is the effective thickness of the spectrum recorded at the lowest temperature (here 80 K). The temperature dependence of $-\ln(f/f_o)$ is presented in Fig. 2. It is evident that three characteristic ranges can be distinguished: (I) for *T ≤ ~145 K* where the dependence is linear i.e. $^{57}$Fe atoms experience harmonic vibrations. The characteristic effective Debye temperature for that range was determined as $\Theta_{eff}$ = 185.0±1.5 K; (IIA) for *~145 K ≤ T ≤ ~240 K* and (IIB) for *~240 K ≤ T ≤ ~350 K* where the dependence is not linear, hence anharmonic effects are important. Following the theory by Maradudin and Flinn, the data from the ranges IIA and IIB were analyzed in terms of equ. (2) yielding $\Theta_{eff}$ =

156.2±13.0 K and $\varepsilon = 18 \cdot 10^{-4}$ K$^{-1}$ for IIA, and $\Theta_{eff} = 151.6±18.0$ K and $\varepsilon = 13.6 \cdot 10^{-4}$ K$^{-1}$ for IIB. The data in range IIB can be alternatively interpreted in terms of two linear sub-ranges; one with ~240 K ≤ T ≤ ~280 K and another with ~280 K ≤ T ≤ ~350 K. The linear fit to the former sub-range yields $\Theta_{eff}$ = 266.4 K and that to the latter one $\Theta_{eff}$ = 501.1 K. Such interpretation would mean that for T ≥ ~240 K a dramatic increase of the lattice hardening has taken place. In addition, a new characteristic temperature of ~280 K should be introduced. Consequently, we think that the interpretation of the behavior found in range IIB in terms of two linear sub-ranges has no physical meaning, hence it will not be further discussed. The $\Theta_{eff}$ – values of 185.0 K, 156.2 K and 151.6 K can be further used to determine relative force constants, $r = \gamma_{Fe-Cr}/\gamma_{Cr-Cr}$. For that purpose the following formula was applied [7]:

$$\Theta_{eff} = (M_{Cr}/M_{Fe})^{1/2}(\gamma_{Fe-Cr}/\gamma_{Cr-Cr})^{1/2}\Theta_D \qquad (3)$$

where $M_{Cr}$ and $M_{Fe}$ are the masses of the host (Cr) and the impurity atom (Fe), while $\gamma_{Fe-Cr}$ and $\gamma_{Cr-Cr}$ are the force constants of the impurity-host and of the host-host binding, respectively. $\Theta_D$ is the Debye temperature of the host (Cr). Thus $r$ = 0.096 was found for range I, $r$ = 0.0673 for range IIA and $r$ = 0.0634 for range IIB. This is 6 – 10 times less than the lowest value known up-to-now (for Fe impurities in Palladium) [8]. The $r$ – values presently found also mean that the coupling of Fe atoms into the Cr-matrix is extremely weak. It is for range I by a factor of ~11 and for ranges IIA and IIB by ~15 weaker than the coupling between the host atoms themselves. On the other hand, the anharmonic effect seems to be unusually strong if one compares presently found $\varepsilon$ - values with those determined for other matrices [2,5].
In an attempt to find an answer what might be the reason for such anomalous dynamics, one has to realize that metallic chromium itself has unusual properties. In particular, its electronic and lattice structure is, below the Néel temperature of ~312 K, harmonically modulated, the phenomenon known as spin-density waves (SDWs), charge-density waves (CDWs) and strain-waves (SWs) [9]. All these three kinds of "waves" are coupled with each other, which certainly makes the interpretation of the results even more difficult. All the more so, there is no adequate theory that could be quantitatively used for this purpose, although there are some theoretical papers pertinent to the issue [10-15]. According to Ref. 10, the effect of phase fluctuations on the lattice dynamics is different in incommensurate structures, such as that observed in chromium, than in normal ones. In particular, they reduce the mean amplitude of atom displacement and produce specially modulated fluctuations of these displacements. In other words, the dynamics of the harmonically distorted lattice itself might give rise to anomalies. In our case, the lattice has not only this kind of distortion, but additionally, it is magnetic and its magnetic moments show harmonically modulated structure (SDWs). Furthermore, in our experiment one observes oscillations of the lattice indirectly i.e. via probe Fe atoms that themselves possess magnetic moment [16]. So there is a fundamental question, whether or not the dynamics of the probe atoms reflects the dynamics of underlying lattice. The question is justified in the light of theoretical calculations on a possible effect of impurities on the SDWs [10-14]. A general conclusion from these calculations is that nonmagnetic impurities pin strongly CDWs while magnetic impurities, such as Fe, do so for SDWs (strongly means that their effect is of the order of a gap, which in the case of chromium is equal to ~0.1eV). In the light of a strong coupling between SDWs, CDWs and SWs, such behavior of the magnetic impurities means that they can also influence the dynamics of the lattice. From experimental viewpoint, the situation seems to be not so clear, and in any case one cannot make a distinct division into nonmagnetic and magnetic impurities [17]. Concerning the former, there are such ones that seem to act as ideal probe nuclei i.e. all features characteristic of the SDWs of a pure chromium can be revealed using them as probes.

Here tantalum [18] and cadmium [19] in PAC experiments and tin in MS measurements [20,21] can be given as best examples. On the other hand, nonmagnetic vanadium behaves like a SDWs killer as its presence quenches the SDWs decreasing $T_N$ at the rate of ~20 K/at%. Magnetic impurities have been revealed to have a strong effect. In particular, iron decreases $T_N$ and simultaneously changes the character of the SDWs from incommensurate to commensurate with a critical concentration of ~2.3 at%. In view of the above described situation, one can merely speculate on the origin of the anomalies in the dynamical effects observed in this study. Below we will try to give some arguments in favor of connecting these effects with the SDWs as well as to list those facts that are against such interpretation. Concerning the former, it should be first recalled that the SDWs exist as two different phases: (i) the high-temperature phase, 123 K ≤ T ≤ 312 K, where the polarization is perpendicular to the wave vector (TSDW), and (ii) the low-temperature phase, T ≤ 123 K, where the polarization is longitudinal (LSDW). It is plausible that the dynamics found in this experiment is connected with these two phases as according to previous studies [22,23], TSDW and LSDW phases have different magnetomechanical properties. In particular, de Morton has observed a decrease in damping of torsional oscillations in the LSDW phase, the effect was predicted by Overhauser [24]. Also the results found by Street gave a clear evidence that the Young modulus and a logarithmic decrement of longitudinal oscillations are characteristic of the two phases [23]. Finally, as illustrated in Fig. 3, the width of the Mössbauer spectra measured at half maximum, that can be regarded as the average amplitude of the SDWs, exhibits two anomalies: one at $T \approx 143$ K, below which the increase of the width is linear, and another one at $T \approx 240$ K where the width has a local maximum. The two temperatures coincide very well with those at which the *f*-factor shows anomalies. However, these arguments and observations do not explain why the dynamics below ~145 K is harmonic and above ~145 K it is not. On the other hand, against a direct relationship between the observed dynamics and the SDWs, the following arguments can be given. First of all, as revealed in this study, the coupling between the probe Fe atoms and the lattice, hence SDWs, is extremely week, so the probability that the dynamics of the former reflects that of the latter is rather low. Secondly, the characteristic temperatures found in the present study i.e. ~145 K, as the border between the harmonic and anharmonic behavior, and ~240 K as the boundary between two anharmonic regions, have no direct correspondence in the SDWs themselves, for which only one characteristic temperature i.e. ~123 K exists and it marks the transition between LSDW and TSDW phases. Thirdly, the temperature of ~145 K which indicates the transition between the harmonic and anharmonic behavior is close to the Kondo temperature for the Cr-Fe system [25], the effect that has nothing to do with the SDWs. In these circumstances one can only hope that the results presented in this Letter will challenge theoreticians to carry out calculations that will shed more light on the dynamics of Fe impurity atoms in chromium.

**Acknowledgements**

The study was supported by the Ministry of Science and Higher Education, Warszawa.

[*] Corresponding author: dubiel@novell.ftj.agh.edu.pl (S. M. Dubiel)

**References**

[1] G. Chandra, C. Bansal and J. Rey, Phys. Stat. Sol. (a), 35, 73 (1976)

[2] W. A. Steyert and R. D. Taylor, Phys. Rev., 134, A716 (1964)

[3] G. P. Gupta and K. C. Lal, Phys. Stat. Sol. (b), 51, 233 (1972)

[4] K. Sørensen and G. Trumpy, Phys. Rev. B, 7, 1791 (1973)

[5] D. G. Howard and R. H. Nussbaum, Phys. Rev. B, 9, 794 (1974)

[6] A. A. Maradudin and P. A. Flinn, Phys. Rev., 129, 2529 (1963)

[7] W. M. Visscher, Phys. Rev., 129, 28 (1963)

[8] D. A. O'Connor, M. W. Reeks and G. Skyrme, J. Phys. F: Metal. Phys., 2, 1179 (1972)

[9] E. Fawcett, Rev. Mod. Phys., 60, 209 (1988)

[10] J. D. Axe, Phys. Rev. B, 21, 4181 (1980)

[11] G. Grüner, Sol. Stat. Phys., 10, 183 (1983)

[12] P. F. Tua and J. Ruvalds, Phys. Rev. B, 32, 4660 (1985)

[13] Ch. Seidel, Phys. Stat. Sol. (b), 149, 327 (1988)

[14] I. Tüttö and A. Zawadowski, Phys. Rev. Letter., 60, 1442 (1988)

[15] G. Grüner, Rev. Mod. Phys., 66, 1 (1994)

[16] B. Babic, F. Kajzar and G. Parette, J. Phys. Chem. Solids, 41, 1303 (1980)

[17] E. Fawcett, H. L. Alberts, V. Yu. Galkin, N. R. Noakes and J. V. Yakhmi, Rev. Mod. Phys., 66, 25 (1994)

[18] G. Teisseron, J. Berthier, P. Peretto, C. Benski, M. Robin, and S. Choulet, J. Magn. Magn. Mater., 8, 157 (1978)

[19] R. Venegas, P. Peretto, G. N. Rao and L. Trabut, Phys. Rev. B, 21, 3851 (1980)

[20] B. Window, J. Phys. C, 2, S210 (1970)

[21] S. M. Dubiel, Phys. Rev. B, 29, 2816 (1984)

[22] M. E. de Morton, Phys. Rev. Lett., 10, 208 (1963)

[23] R. Street, Phys. Rev. Lett., 10, 210 (1963)

[24] A. W. Overhauser, Phys. Rev., 128, 1437 (1966)

[25] S. Katano and N. Mori, J. Phys. Soc. Jpn., 49, 1812 (1980)

**Figures**

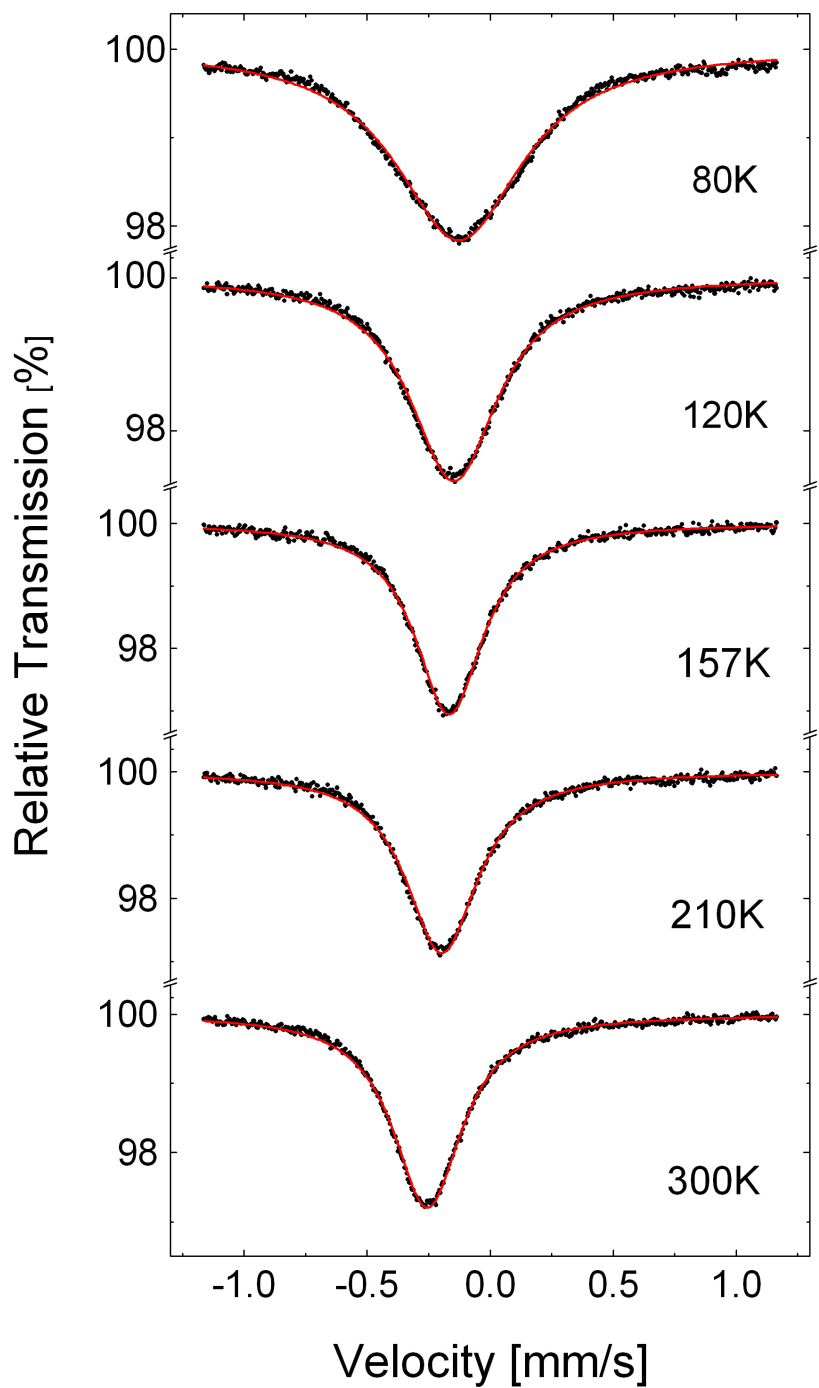

Fig. 1

$^{57}$Fe site Mössbauer spectra recorded at various temperatures shown. The solid lines are the best-fits obtained with the procedure as described in the text.

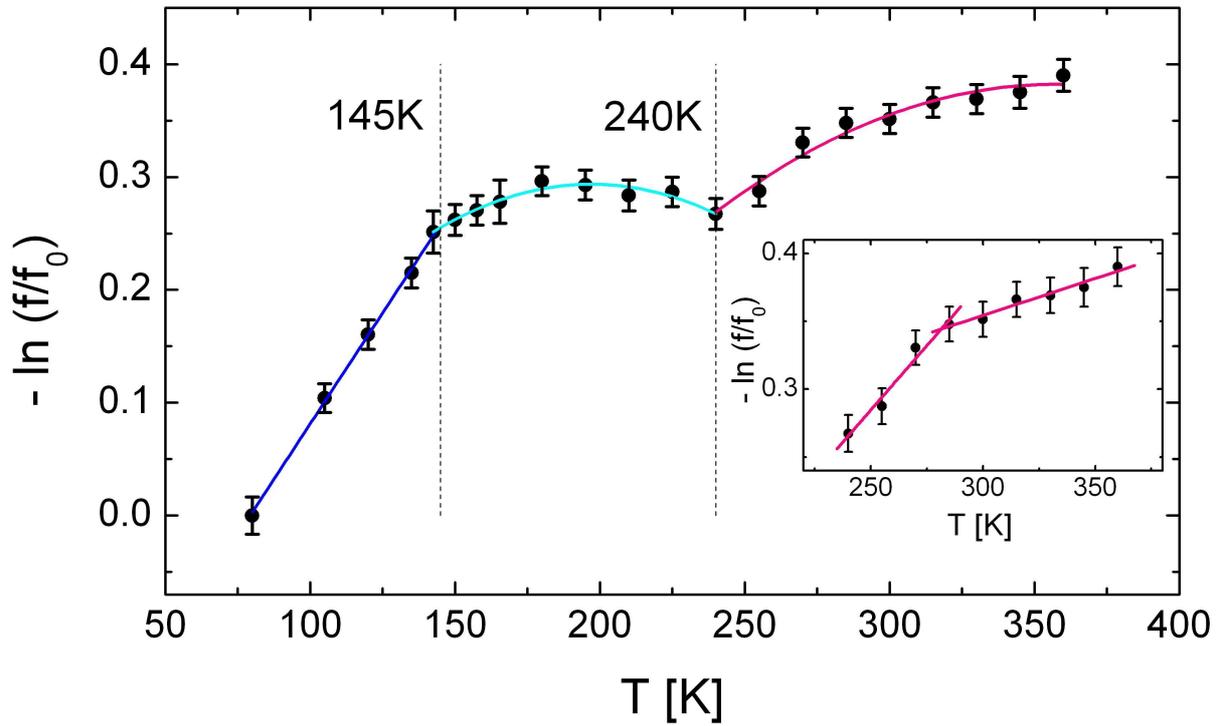

Fig. 2

(Online color) Logarithm of the relative *f*-factor versus temperature. The solid lines represent the best fits to the data in three temperature ranges in terms of equ. (2). The vertical dashed lines indicate the borders between the three ranges. The inset shows an alternative fit to the data from the range IIB in terms of two straight lines.

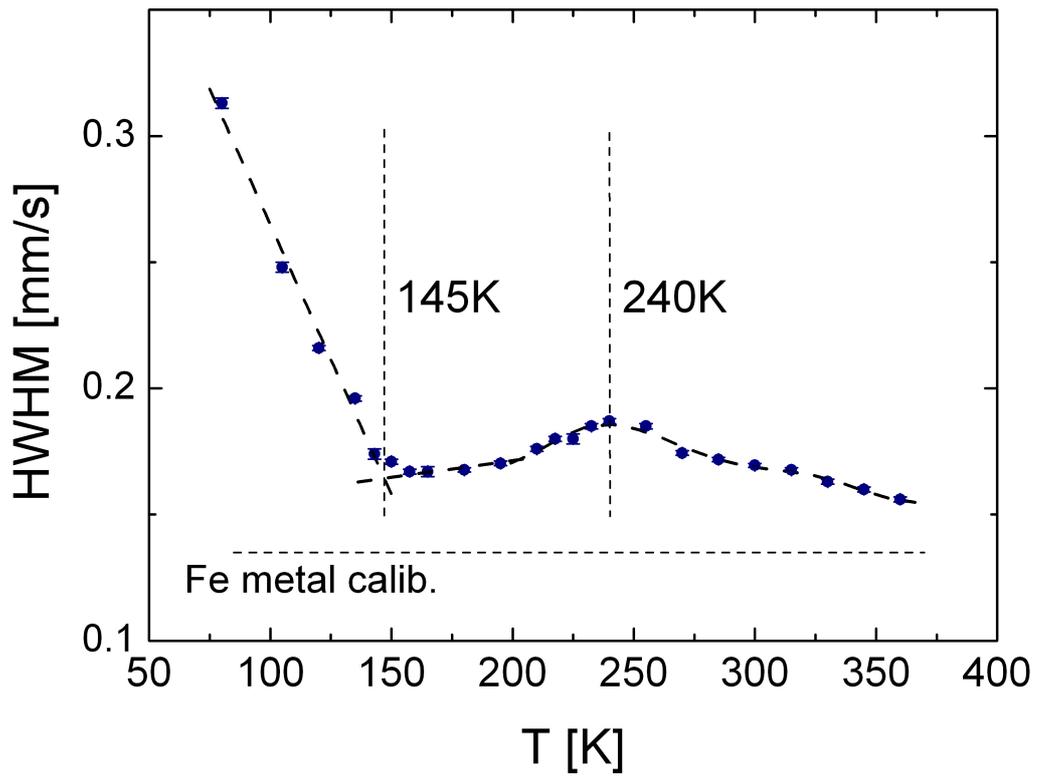

Fig. 3

(Online color) Half width at half maximum versus temperature as derived from the measured spectra. The vertical dashed lines mark anomalies i.e. the border between linear and nonlinear behavior (145 K) and a local maximum (240 K).